# Friendship Paradox and Attention Economics

Subhash Kak[1]

Abstract. The friendship paradox is revisited by considering both local and global averages of friends. How the economics of attention affects the recruitment of friends is examined. Statistical implications of varying individual attentions are investigated and it is argued that this is one reason why the mean of friends is higher than the median in social networks. The distribution of friends skews to the right for two other reasons: (i) the presence of institutional nodes that increase the mean; and (ii) the dormancy of many of the nodes. The difference between friends and friends of friends is a measure of the structural information about the network.

1. Introduction

According to the friendship paradox "our friends have on average more friends than we have" [1]. It provides insights in a variety of situations in social and information networks. It was shown, for example, that immunization of random acquaintances of random nodes (individuals) leads to a reduction of immunization threshold [2]. The paradox has also applicability in citation networks [3].

The implication of the normal statement of the paradox is that one is speaking of the friends one knows, which means we are speaking of the local mean of friends. The mean number of friends can also be computed for the entire membership of the graph and this is likely to be different from the varying values of the local mean.

This mathematical explanation that is usually provided to the friendship paradox is in terms of the global mean of friends. It should be noted that the knowledge members possess about the number of their friends affects their behavior. The matter of awareness brings up the question of economy of attention [4]-[7]. In the age of computers it is not information that is scarce but rather the attention that can be paid to it by individuals. Therefore in a social circle the behavior to approve someone else's request to be a friend is related to the amount of attention available to this individual. The desire to recruit as many friends as possible for bragging rights faces the reality of the economics of attention. This leads to a game theoretic situation of the kind that has provided insight on the behavior of actors in a social network [7]. There is additionally the desire for anonymity [8], which in the past was expressed in the use of masks and now in the use of avatars.

One should also note Dunbar's number [9] (and others similarly determined measures) according to which the number of stable relationships (or friends) a person can have is around 150. According to a Facebook researcher [10],[11], the average number of friends, μ, on Facebook in 2011 stood at 190 whereas the median, M, was about 100, indicating that a large number of friends have a low number of friends. In the Facebook distribution, the degree (number of friends) ranges from 1 to 5,000, the higher number is the limit that was imposed by the company. The data showed short average path lengths and high clustering geographically and with respect to age. The fact that the highest number of friends in the network stood in the thousands is also because the network consists not only of individuals but also corporations and businesses whose count of friends has nothing to do with the cognitive limits of the kind described by Dunbar.

In this article, we look at different aspects of the friendship paradox related to the economics of attention. We argue that another reason that contributes to the mean of friends being higher than the median in social networks is the high connectivity of institutional entities that have degree running into the thousands.

---
[1] School of Electrical and Computer Eng., Oklahoma State University, Stillwater, OK 74078; subhash.kak@okstate.edu

## 2. Global and Local Averages of Friends of Friends (FF)

A social network is a graph. The nodes are the individual members and the edges represent connectivity. The number of connections from a node to all other nodes represents the number of links (or friends) of that node.

Let the total number of nodes in a network be *n*. Let the edges connected to node *i* be $x_i$, which is the number of friends *i* has, and it may be represented also by $F(i)$. Also let the connectivity between the nodes *i* and *j* be represented by $x_{ij}$ which is a binary variable being 0 if there is no connection and 1 if the two nodes are connected. First note, that

$$x_j = \sum_i x_{ij} \tag{1}$$

Furthermore,

$$\sum_i x_i^2 = \sum_i \sum_j x_i x_{ij} \tag{2}$$

The number of friends of friends (FF) of node *i* is determined by counting the friends of friends of *i*.

Also note that the friends of friends of node i equals:

$$FF(i) = \sum_j x_{ij} x_j \tag{3}$$

There are now two means of friends of friends that we call $\mu_{FF}(i)$ and $\mu_{FF}(global)$:

$$\mu_{FF}(i) = \frac{FF(i)}{x_i} \tag{4}$$

$$\mu_{FF}(i) = \frac{\sum_j x_{ij} x_j}{\sum_j x_{ij}} \tag{5}$$

$$\mu_{FF}(global) = \frac{\sum_i x_i^2}{\sum_i x_i} \tag{6}$$

The mean number of edges (or friends) is $\mu_F$:

$$\mu_F = \frac{1}{n} \sum_i x_i \tag{7}$$

The total number of friends of friends is given by $\sum_i x_i^2$. We can also write that

$$VAR(X) = \sigma_X^2 = \frac{1}{n}\sum_i x_i^2 - \mu_F^2 \tag{8}$$

$$\sum_i x_i^2 = n\mu_F^2 + n\sigma^2 \tag{9}$$

$$\frac{\sum_i x_i^2}{\sum_i x_i} = \mu_{FF} = \frac{n\mu_F^2 + n\sigma^2}{n\mu_F} \tag{10}$$

$$\mu_{FF} = \mu_F + \frac{\sigma^2}{\mu_F} \tag{10}$$

In other words,

$$\mu_{FF} \geq \mu_F \tag{11}$$

This is the purported proof of the friendship paradox.

It follows that for a fully connected network of *n* nodes, for which σ² is zero, $\mu_{FF} = \mu_F$ and it equals *n-1*. Likewise for other networks where the connectivity is the same for each node, as in the network where each node has a single link (as in 1 and 2 are connected, and 3 and 4, and so on), the two are equal. Since equality is obtained when the network is regularly connected, we can conclude that the level of inequality in (11) depends on the nature of connectivity of the networks. One might propose that the level of inequality (which is the variance associated with the degree) codes the amount of information in the structure of the network. Information theoretic approaches to the information in the structure of the network may also be adopted [12],[13].

The result (11) does not address the information related to friends at the local level. What we ought to be looking for is that on an average the mean of the $\mu_{FF}(i)$ is greater than $x_i$.

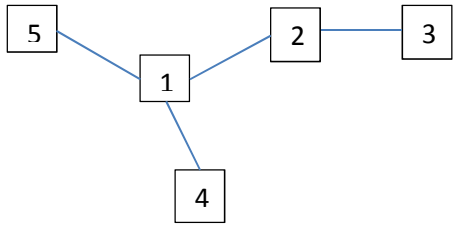

Figure 1. Network 1

*Example 1.* Table 1 gives the number of friends (F) for the network of Figure 1 that consists of five nodes 1, 2, 3, 4, 5; it also gives the number of friends of friends (FF).

Table 1. Connectivity of the network of Figure 1 ($\mu_{FF}(global)=2$)

| Node | F(i) | FF(i) | $\mu_{FF}(i)$ | $\mu_{FF}(i) >$ F(i) |
|---|---|---|---|---|
| 1 | 3 | 4 | 1.33 | no |
| 2 | 2 | 4 | 2.00 | - |
| 3 | 1 | 2 | 2.00 | yes |
| 4 | 1 | 3 | 2.00 | yes |
| 5 | 1 | 3 | 2.00 | yes |

In this example, out of 5 cases, three have more friends than their own and one has value that is equal; only one case (node 1) has friends more than that of friends. As is clear, this is due to the double counting of edges in the definition of friends.

Various nodes are not likely to know the global value of *FF*. What they know is the average value of friends of their friends which is *FF(i)*.

*Example 2.* Consider the network of Figure 3. It consists of 5 nodes

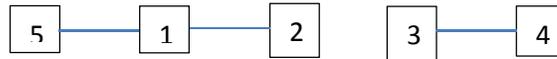

Figure 2. Network 2

Table 2. Connectivity of the network of Figure 2 ($\mu_{FF}(global)=1.33$)

| Node | F(i) | FF(i) | $\mu_{FF}(i)$ | $\mu_{FF}(i) >$F(i) |
|---|---|---|---|---|
| 1 | 2 | 2 | 1.00 | no |
| 2 | 1 | 2 | 2.00 | yes |
| 3 | 1 | 1 | 1.00 | no |
| 4 | 1 | 1 | 1.00 | no |
| 5 | 1 | 2 | 2.00 | yes |

The $\mu_{FF}(global)$ is larger than local count of friends in 4 out of 5 cases. On the other hand, the local mean of friends of friends is greater than local friends in only 2 out of 5 cases; in 2 cases it is identical; and in one case it is less.

In other words, we see that the local count of friends of friends as compared to friends may not be as different as in the global case of inequality (11).

*Example 3.* Consider the network of Figure 4. It consists of 5 nodes

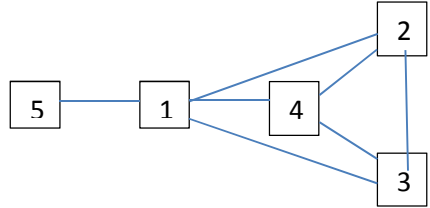

Figure 3. Network 3

Table 3. Connectivity of the network of Figure 4 ($\mu_{FF}(global)=3.14$)

| Node | F(i) | FF(i) | $\mu_{FF}(i)$ | $\mu_{FF}(i)$ >F(i) |
|------|------|-------|---------------|---------------------|
| 1 | 4 | 10 | 2.50 | no |
| 2 | 3 | 10 | 3.33 | yes |
| 3 | 3 | 10 | 3.33 | yes |
| 4 | 3 | 10 | 3.33 | yes |
| 5 | 1 | 4 | 4.00 | yes |

In this example we see that for 80 percent of nodes the number of friends of friends is larger than the number of friends.

## 3. Economy of Attention

The attention one can pay to friends decreases as their number increases. At the same time the amount of information that is available to a user increases as the number of friends increases. The first of these is like the demand curve whereas the amount of information is supply.

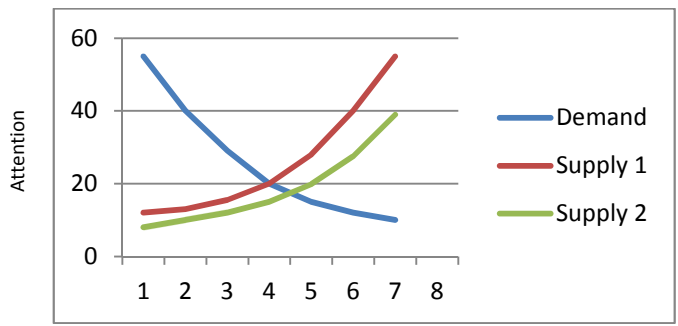

Figure 4. Supply and demand curves for friends
(x-axis: count of friends; y-axis: cost)

In Figure 4, the demand curve represents attention and the x-axis represents the count of friends. When the number of friends is low, the individual can allocate more attention to each than when the number of friends is high. Hence the negative slope of this curve. On the other hand, the supply of information from the friends increases as the number of friends increases (Supply 1). Like the supply-demand curves of traditional economics, the point at which the two curves intersect represents the equilibrium point. If the supply from each friend goes down (Supply 2), the number of friends at the equilibrium point will be greater.

As mentioned in the Introduction, not all nodes in a social network are individuals. There are many that are institutions, groups, or businesses. Such nodes, for which we will use the collective label of *institution*, do not have the same constraints of attention since these nodes serve as clearing houses of information or as chat sites. Such institutional sites will have many, many more links than those that represent real individuals.

Institutional nodes skew the distribution of friends towards the right. Furthermore, the fact that most users are relatively inactive shifts the supply curve towards the right that further skews the distribution. Both of these contribute to the mean becoming much greater than the median as is true for Facebook data [10].

### 4. Relationship between Median and Mean

Let the median and mean of friends in a social network be denoted by M and μ, respectively. As mentioned before, the values for these in 2010 were 100 and 190. This tells us [14] that the standard deviation σ is at least 90.

We take the number of friends to be random variable X. Since the absolute-value and the square functions are convex and the median minimizes the absolute deviation function:

$$|\mu - M| = |E(X - M)| \leq E(|X - M|) \tag{12}$$

$$\leq E(|X - \mu|) \leq \sqrt{E((X - \mu)^2)} = \sigma \tag{13}$$

Note further the data [10] indicates that the distribution of friends is essentially unimodal. Therefore, the bound (13) can be made tighter [15].

The degree distribution in a random network [16] will be binomial (or Poisson if the nodes is very large) but social networks are not totally random. The skewness of the degree distribution has been studied by many authors (e.g.[17]).

For a simple analysis, the network may be replaced by an aggregate model where the variable takes just a few values but provides the correct median and mean values. Thus the Facebook data may be replaced by, say, 7-point or 9-point distributions as follows:

7-point: (1, 5, 20, 100, 200, 200, 900) with M=100, μ≈ 200, and σ ≈ 375 (14)

9-point: (1, 2, 5, 20, 100, 300, 300, 300, 800) with M= 100, μ≈ 200, and σ ≈ 290 (15)

The measured value of the standard deviation and other modes in the degree distribution curve can be used to pick the specific model.

Let $\mu - M = a$ and we use the knowledge that $a$ is positive and the distribution is skewed to the right. We can further estimate the probability that the number of friends is within the interval $(\mu-a, \mu+a)$. Let $a = \alpha\sigma$, then by Chebyshev's inequality,

$$\Pr(|X - \mu| \geq \alpha\sigma) \leq \frac{1}{\alpha^2} \tag{16}$$

Since $\Pr(X \leq M) = 0.5$, the above inequality can be modified to:

$$\Pr(X - \mu \geq \alpha\sigma) \leq \frac{1}{\alpha^2} - 0.5 \tag{17}$$

This is equivalent to:

$$\Pr(X \geq 2\mu - M) \leq \frac{\sigma^2}{(\mu - M)^2} - 0.5 \tag{18}$$

This inequality can be useful for distributions where $\alpha$ is close to 1. For the specific case when it is 1, the inequality may be rewritten as:

$$\Pr(X \geq \mu + \sigma) \leq 0.5 \tag{19}$$

But, of course, such a case will not be typical of social networks as the value of $\sigma$ for these is likely to be much higher than the difference between the mean and the median.

Since social networks are evolving networks, the characteristics will keep on changing under pressures from other competing networks and also the changing behavior of participants. The behavior of the network is likely to have a chaotic evolutionary component [18].

### 5. Conclusions

This paper has shown how the averages for friends of friends on the one hand and friends on the other can vary greatly. The friendship paradox was described in terms of both local and global averages of friends. How the economics of attention affects the recruitment of friends was discussed. It was shown that the distribution of friends skews to the right for three reasons: (i) constraints on attention; (ii) the presence of institutional nodes that increases the mean; and (iii) the shifting of the equilibrium to the right due to the dormancy of many of the nodes.